\documentclass[useAMS,usenatbib,usegraphicx]{mn2e}
\usepackage[a4paper,centering, totalwidth=520pt, totalheight=700pt]{geometry}
\usepackage{hyperref}
\usepackage{subfig}
\usepackage{graphicx}
\usepackage{color}
\usepackage{pdfpages}
\usepackage{amsmath}
\usepackage{float}

\newcommand{\be}{\begin{equation}}
\newcommand{\ee}{\end{equation}}

\def\ltsima{$\; \buildrel < \over \sim \;$}
\def\simlt{\lower.5ex\hbox{\ltsima}}
\def\gtsima{$\; \buildrel > \over \sim \;$}
\def\simgt{\lower.5ex\hbox{\gtsima}}
\def\la{$\langle$}

\title{Einasto Profiles and the Dark Matter Power Spectrum}

\author[Ludlow \& Angulo] {\parbox{18cm}{
    Aaron D. Ludlow$^{1,\star}$ \& Ra\'ul E. Angulo$^{2}$}\vspace{0.3cm}\\
$^{1}${Institute for Computational Cosmology, Dept. of Physics, Univ. of
  Durham, South Road, Durham  DH1 3LE, UK}\\
$^{2}${Centro de Estudios de F\'isica del Cosmos de Arag\'on,
  Plaza San Juan 1, Planta-2, 44001, Teruel, Spain}\\}

\begin{document}

\vspace{-1cm}
\maketitle 

\begin{abstract}
  We study the mass accretion histories (MAHs) and density profiles of dark matter halos 
  using N-body simulations of self-similar gravitational clustering from scale-free
  power spectra, $P(k)\propto k^n$. 
  We pay particular attention to the density profile {\em curvature}, which we 
  characterize using the shape parameter, $\alpha$, of an Einasto profile. 
  In agreement with previous findings our results suggest that, despite vast 
  differences in their MAHs, the density profiles of virialized halos are remarkably 
  alike. Nonetheless, clear departures from self-similarity are evident: 
  for a given spectral index, $\alpha$ increases slightly but
  systematically with ``peak height'', $\nu\equiv\delta_{sc}/\sigma(M,z)$,
  regardless of mass or redshift. More importantly, however, the ``$\alpha-\nu$''
  relation depends on $n$: the steeper the initial power spectrum, the more gradual
  the curvature of both the mean MAHs and mean density profiles. These results are consistent with 
  previous findings connecting the shapes of halo mass profiles and 
  MAHs and imply that dark matter halos {\em are not} structurally self-similar but, 
  through the merger history, retain a memory of the linear density field from which they form. 
\end{abstract}


\renewcommand{\thefootnote}{\fnsymbol{footnote}}
\footnotetext[1]{E-mail: aaron.ludlow@durham.ac.uk} 

\section{Introduction}
\label{SecIntro}

The density profiles of dark matter (DM) haloes are
well described by the Einasto (1965) \nocite{Einasto1965} profile:
\begin{equation}
  \ln \, (\rho_{\rm E}(r)/\rho_{-2})=-2/\alpha\,[(r/r_{-2})^\alpha-1].
  \label{eq:rho_einasto}
\end{equation}
Here $\rho_{-2}$ and $r_{-2}$ are characteristic values density and 
radius, and $\alpha$ is a ``shape'' parameter that governs the 
profile's curvature. The scaling parameters -- commonly
cast in terms of virial mass\footnote{We define the virial mass, ${\rm M}_{200}$,
  as that enclosed by a sphere of mean density $200\times\rho_{\rm crit}$
  surrounding the halo particle with the minimum potential energy. This
  implicitly defines the virial radius as ${\rm M}_{200}=(800/3)\pi\rho_{\rm crit} r_{200}^3$.}, 
${\rm M}_{200}$, and concentration, $c\equiv r_{200}/r_{-2}$ -- are not 
independent, but correlate in a way that encodes the formation
history-dependence of halo structure: halos collapsing early, when the 
Universe was dense, inherit higher characteristic densities (or concentrations) 
than those collapsing later \citep[e.g.][]{Navarro1997}. This 
idea led to the development of a number of analytic and empirical models 
that successfully describe the mass, redshift, cosmology and power spectrum 
dependence of halo concentrations \citep[e.g.][]{Maccio2008,Diemer2015,Ludlow2016}. 

Compared to concentration the shape parameter, $\alpha$, has received little
theoretical attention. Its weak but systematic dependence on halo mass and redshift 
reported by \cite{Gao2008} disclosed a simpler underlying relation when expressed 
in terms of dimensionless peak height\footnote{The ``peak height'' is a
dimensionless mass variable defined $\nu({\rm M},z)=\delta_{\rm
  sc}/\sigma({\rm M},z)$, where $\delta_{\rm sc}$ is the spherical
top-hat collapse threshold and $\sigma({\rm M},z)$ is the rms mass
fluctuation in spheres of mass ${\rm M}$. Note that $\nu({\rm M_{NL}},z)=1$ 
defines the non-linear mass scale, ${\rm M_{NL}}$.}, $\nu$. On average, $\alpha\approx 0.16$ for 
all $\nu\simlt 1$, but increases to $\sim$0.3 for the rarest halos in their 
simulations. This result has been supported by a number of subsequent studies 
\citep[e.g.][]{Dutton2014} but the physical origin 
of the relation has not been pinned down.

Nevertheless, the need for a third parameter is clear and plausible
interpretations for its origin have been put forth. \citet{Ludlow2013} suggested that, when 
expressed in appropriate units, the shape of the {\em average} halo mass profile 
is the same as that of the {\em average} MAH: both are approximately
universal and well described by an Einasto 
profile with the {\em same} shape parameter ($\alpha\approx 0.18$; see \cite{Ludlow2016}). 
Intriguingly, halos whose MAHs deviate in a particular way from the mean have 
mass profiles that deviate from the mean in a similar way suggesting
that, at fixed mass, halos that assemble more rapidly than average
exhibit more ``curved'' mass profiles, and vice versa. The
correlation is weak, however, and substantial deviations from the
mean MAH are required to leave a noticeable imprint on $\alpha$.

\citet{Cen2014} argued that profiles similar to eq.~\ref{eq:rho_einasto} may be a 
natural outcome of gravitational clustering in models seeded by Gaussian density 
fluctuations. He conjectured that centrally concentrated halos with extended outer 
envelopes (corresponding to small values of $\alpha$) form primarily through mergers 
of many small, dense clumps; the expected outcome for power spectra dominated by short wavelength modes. 
Conversely, smooth and coherent collapse occurs when the fluctuation field is 
dominated by long wavelength modes, due to the lack of significant substructure. In this 
case, diffuse accretion plays a vital role in halo growth and the resulting profile is 
shallow in the center and steep in the outskirts. This hypothesis is backed-up by  
numerical experiments \citep{Nipoti2015}. 

Finally, \citet{Angulo2016} showed that a rapid 
succession of major mergers leads to a remnant whose density profile is more curved than 
that of its progenitors, supporting the idea that halo profiles are sensitive to the 
precise details of how their mass was assembled. 

What determines the shapes of halo mass profiles? The answer will 
illuminate the processes that establish the structural 
properties of DM halos, and may lead to improvements to future models for 
halo structure. We address this issue here using a suite of self-similar simulations 
of gravitational clustering. Our simulations and their analysis are described in Section~\ref{SecSimsAnal}; 
the MAHs and density profiles of halos in each are presented in Sections~\ref{SecMAH} 
and \ref{SecRho}, respectively. Section \ref{SecConc} summarizes our findings and provides some 
concluding remarks.

\vspace{-0.5cm}
\section{Simulations and Analysis}
\label{SecSimsAnal}

\subsection{Scale-Free Models}
\label{SecSFSims}

We consider a suite of Einstein-de Sitter models (matter density $\Omega_{\rm
  M}(a)\equiv\rho_{\rm M}(a)/\rho_{\rm crit}(a)=1$) with self-similar power
spectra, $P(k)\propto k^n$, and a scale factor that is a
power-law of time, $a(t)\propto t^{2/3}$.
The only physical scale in such a model is the one at which fluctuations 
become non-linear at a particular time, which is defined by
$\delta_{sc}=\sigma({\rm M_{NL}},a)$, where ${\rm M_{NL}}\propto a^{6/(3+n)}$ 
is the non-linear mass. All of our simulations adopt such a model, but 
change the balance of power between large and small scales by varying the spectral 
index, $n$. For larger values of $n$, the density field is increasingly 
dominated by short-wavelength modes, and the characteristic mass grows very 
slowly. As we will see in Section~\ref{SecMAH}, some control over the growth 
histories of halos can therefore be attained by varying $n$ appropriately. 

We simulated four scale-free models ($n=0$, $-1$, $-2$ and $-2.5$) using $1024^3$ 
particles to evolve the DM. Gravitational forces were softened at a 
fraction $f=0.05$ of the mean inter-particle separation. Although
arbitrary, we set the box size to be  
$L=100\, h^{-1}{\rm Mpc}$ and normalize the power spectrum so that $\sigma_8= 1$
when linearly extrapolated to $a=1$ ($\sigma_8$ is the rms mass fluctuation in 8 $h^{-1}$ 
Mpc spheres). Starting redshifts were chosen to ensure that particle-scale 
fluctuations were safely in the linear regime, at which point
positions and velocities were generated according to second order Lagrangian perturbation theory 
\citep{Jenkins2013} using a different white noise field for each $n$.
Our simulations were carried out with a lean version of the \textsc{gadget} 
code \citep{Springel2005b}. 

The simulations were evolved for a range of expansion factors ending either at 
$a_f=1$, or the most recent time at which $\sigma({\rm M}_{box},a_f)\simlt 0.2$ 
(${\rm M}_{box}$ is the box mass).
This ensures that fluctuations on the box scale remain close to linear at 
the final time, limiting the impact of missing large-scale modes. Our simulations probe a 
very different range of expansion factor: $a_f/a_i\approx 10^4$, $10^3$,
128 and 46.9 for $n=0$, $-1$, $-2$ and $-2.5$, respectively. For each run 
65 snapshots were stored in equally-spaced steps of $\log \, a$, 
with the first output corresponding to the time at which ${\rm
  M_{NL}}$ was equivalent to that of $\sim 20$ particles. Our output sequence
ensures that ${\rm vM_{NL}}$ grows by a constant factor
$\Delta\log {\rm M}=6/(n+3) \Delta \log a$ between snapshots. 

%__________________________________________
\begin{figure}
  \includegraphics[width=0.47\textwidth]{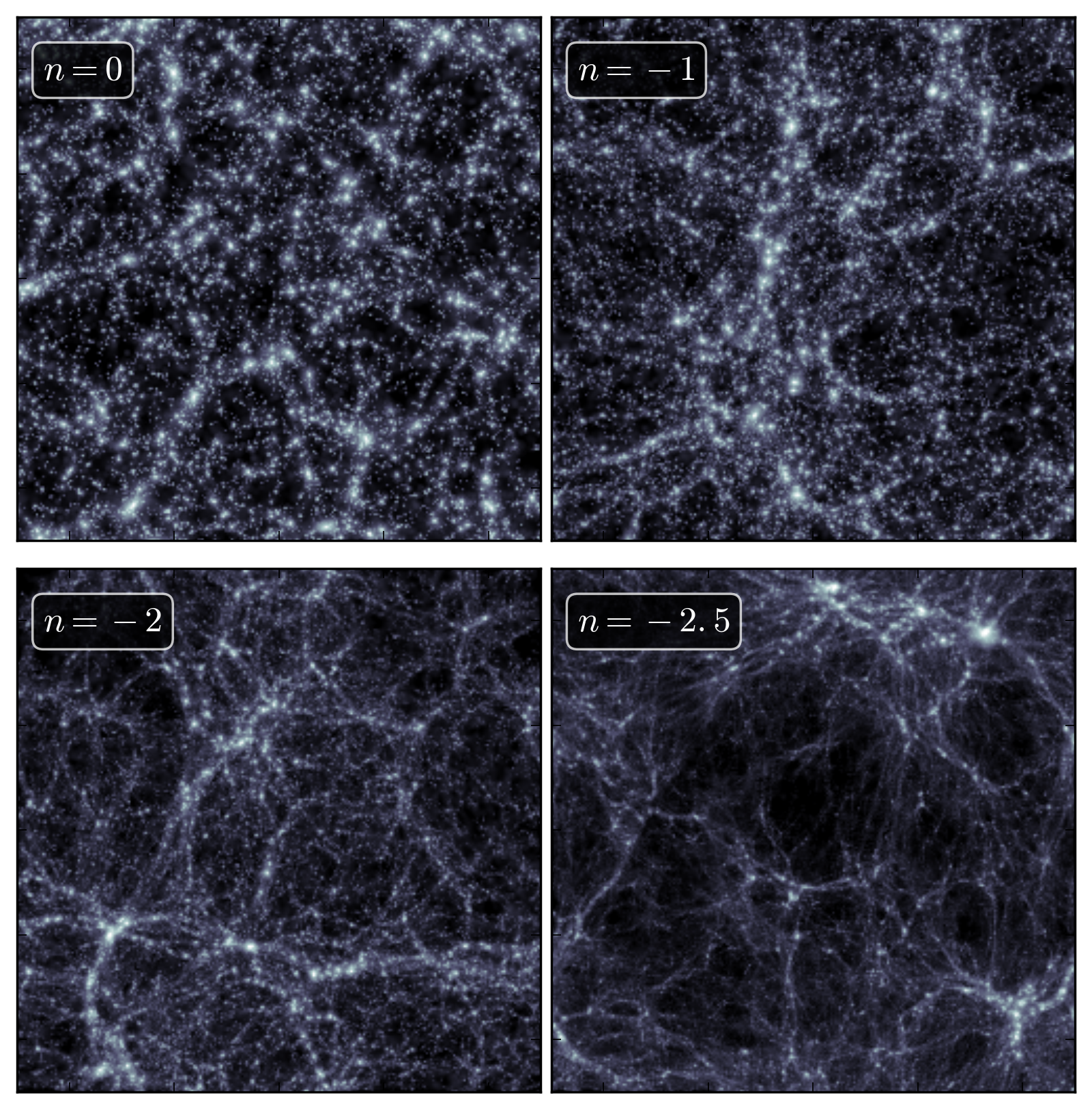}
  \caption{Dark matter distribution in the final output of each of our
  scale-free simulations. All models adopt a scale-invariant linear power spectrum,
  $P(k)\propto k^n$, with spectral indices $n=0$, $-1$, $-2$ and $-2.5$.
  Note that short wavelength modes dominate structure formation for the 
  $n=0$ model (upper left panel), with large scale modes becoming increasingly 
  important as $n$ decreases.}
  \label{fig:ScaleFreeImage}
\end{figure}
%__________________________________________

\begin{figure*}
  \subfloat{\includegraphics[width=0.18\textwidth]{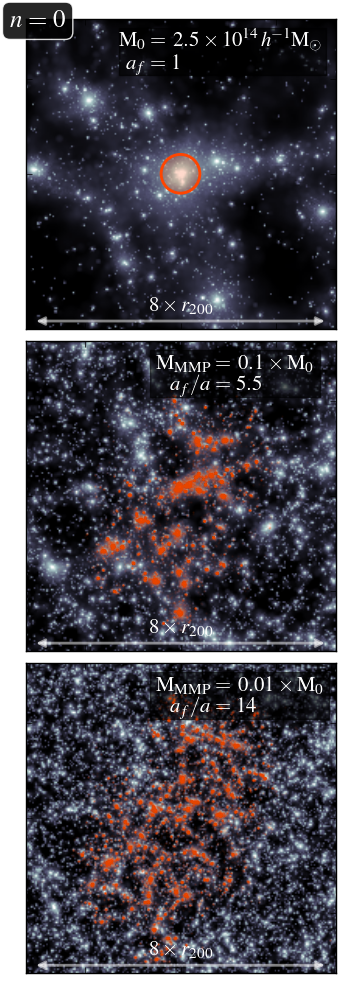}}
  \subfloat{\includegraphics[width=0.18\textwidth]{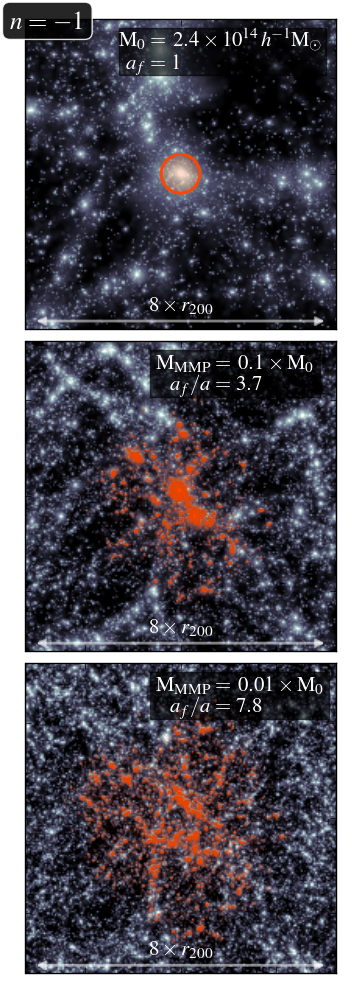}}
  \subfloat{\includegraphics[width=0.18\textwidth]{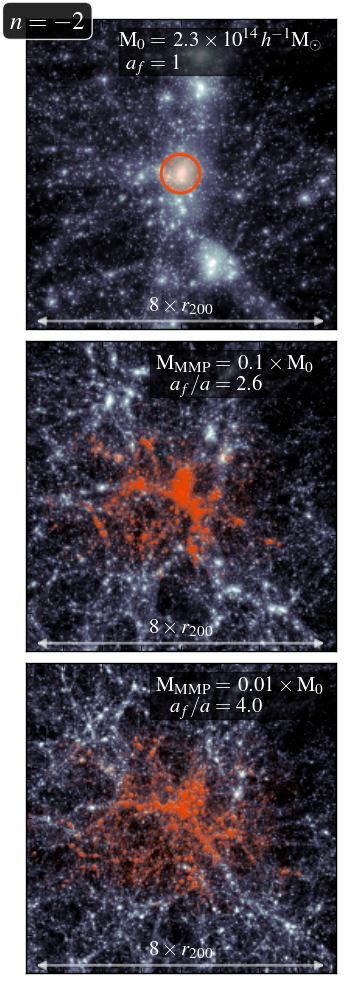}}
  \subfloat{\includegraphics[width=0.18\textwidth]{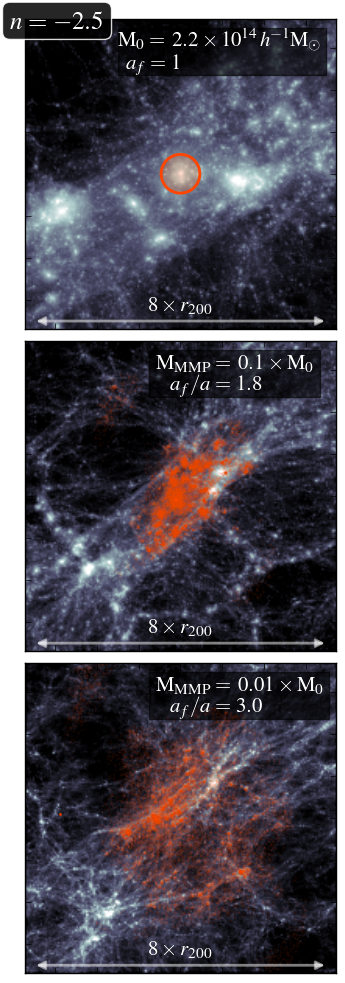}} 
  \caption{The formation of a massive cluster in each of
    our scale-free simulations. From left to right, columns show results for 
    different power-law power spectra, ranging from $n=0$ (left-most)
    to $n=-2.5$ (right-most). Different rows
    correspond to different times during the halo's evolution. 
    Top panels show the final distribution of DM in the halo vicinity; 
    the middle and bottom panels show, respectively, 
    the DM in a region surrounding the halo's main progenitor at 
    the time its virial mass was $\sim 10\%$ and $\sim 1\%$ of $M_0$. 
    In each case, the box-length is fixed to $8\times r_{200}(a_f)$ in comoving units, where 
    $r_{200}(a_f)$ is the virial radius of the halo at the final time.
    The thick orange circle in the upper panels marks $r_{200}$.
    Orange points in other panels highlight the subset of particles that,
    by $a_f$, end up within $r_{200}$ of the halo's descendant.}
  \label{fig:MassiveEvol}
\end{figure*}

Figure~\ref{fig:ScaleFreeImage} provides a visual impression of the DM
distribution in the final output of each simulation. The $n=0$ run is characterized 
by a large number of dense clumps whose large-scale distribution is close to uniform 
across the box. As $n$ decreases, large scale 
modes have a more noticeable impact on the flow of DM and 
prominent features of large-scale structure emerge, such as
voids, filaments and rare clusters.

\subsection{Analysis}
\label{SecAnal}

Friends-of-friends (FoF) halos and their associated 
substructure were identified using \textsc{subfind} \citep{Springel2001b} in
all simulation outputs. The halo catalogs were combined into merger trees 
using the method described in \citet{Jiang2014}, which were then used to build MAHs 
by tracking each halo's main progenitor back through previous simulation 
outputs. We also compute the ``collapsed mass history'' (CMH), defined as the 
{\em total} mass of progenitors larger than $f=10^{-3}$ times the present-day mass, ${\rm M_0}$.
In addition to MAHs and CMHs, we compute two equilibrium diagnostics: 1) the center-of-mass offset, 
$d_{\rm off}=|\mathbf{r}_p-\mathbf{r}_{\rm CM}|/r_{200}$, defined as
the distance between the halo's center-of-mass and most-bound
particle, and 2) the substructure mass fraction\footnote{When computing
  $f_{\rm sub}$ we only consider subhalos whose masses
are at least 1\% of their host's virial mass. 
Our limit of ${\rm N_{200}}\geq 5\times 10^4$ particles thus 
ensures that the lowest mass
subhalos contributing to $f_{\rm sub}$ are resolved with $\simgt$500 particles.}, 
$f_{\rm sub}=M_{\rm sub}(<r_{200})/{\rm M}_{200}$. 
In the remainder of the paper we
will only consider ``relaxed'' halos, defined as those that satisfy both $d_{\rm off}<0.1$
and $f_{\rm sub}<0.1$, and impose a minimum particle number of ${\rm N_{200}}\geq 5\times 10^4$.

We construct the spherically-averaged density profiles for these halos 
in 50 equally-spaced steps in $\log r/r_{200}$ spanning the range $-2.5$
to 0.176 ($r_{\rm max}\approx 1.5\times r_{200}$). These 
profiles are then fit with eq.~\ref{eq:rho_einasto} in order to determine the best-fit values of 
$r_{-2}$ and $\alpha$. Fits are restricted to the radial range $(r_{\rm min}, r_{200})$,
where $r_{\rm min}$ is the larger of $0.02\times r_{200}$ or $2\times\epsilon$.
Best-fit models are obtained by simultaneously adjusting the three parameters
of eq.~\ref{eq:rho_einasto} in order to minimize a figure-of-merit function, defined
\begin{equation}
  \psi^2 = \frac{1}{{\rm N}_{\rm bin}} \sum_{i=1}^{\rm Nbin} [\ln \rho_i - \ln \rho_E(\rho_{-2};r_{-2};\alpha)]^2.  
  \label{eq:fom}
\end{equation}
We obtain best-fit parameters for {\em individual} halos, as well as for {\em median}
profiles in logarithmically-spaced bins of $\nu$.

\vspace{-0.5cm}
\section{Mass Accretion Histories}
\label{SecMAH}

As mentioned above, the rate of clustering in self-similar models 
depends sensitively on $n$. This can be readily seen in 
Figure~\ref{fig:MassiveEvol}, where we plot the growth history of
a massive cluster in each model. From left to right, columns correspond to $n=0$,
$-1$, $-2$ and $-2.5$, respectively; rows to the final simulation output (top), and to 
those at which the halo's main progenitor first reached $\sim$10\% 
(middle) and $\sim$1\% (bottom) of its final mass 
In all cases the halo is resolved at
the final time by $\sim 10^6$ particles within $r_{200}$, which is marked
with an orange circle in the upper panels; all particles within
$r_{200}$ at $z=0$ are highlighted using orange points in other panels.

Halos in different models form differently, and occupy
distinct large-scale environments at the simulation's end point. 
For $n=0$, large quantities of DM have 
assembled into high-density clumps at very early times and 
structures form leisurely, through the slow merging of
many lower mass halos. As $n$ decreases the well-structured pattern of
progenitors loosens and the main clump forms rapidly by
aggregating a number of lower-mass progenitors and diffuse material.

\begin{figure}
  \includegraphics[width=0.48\textwidth]{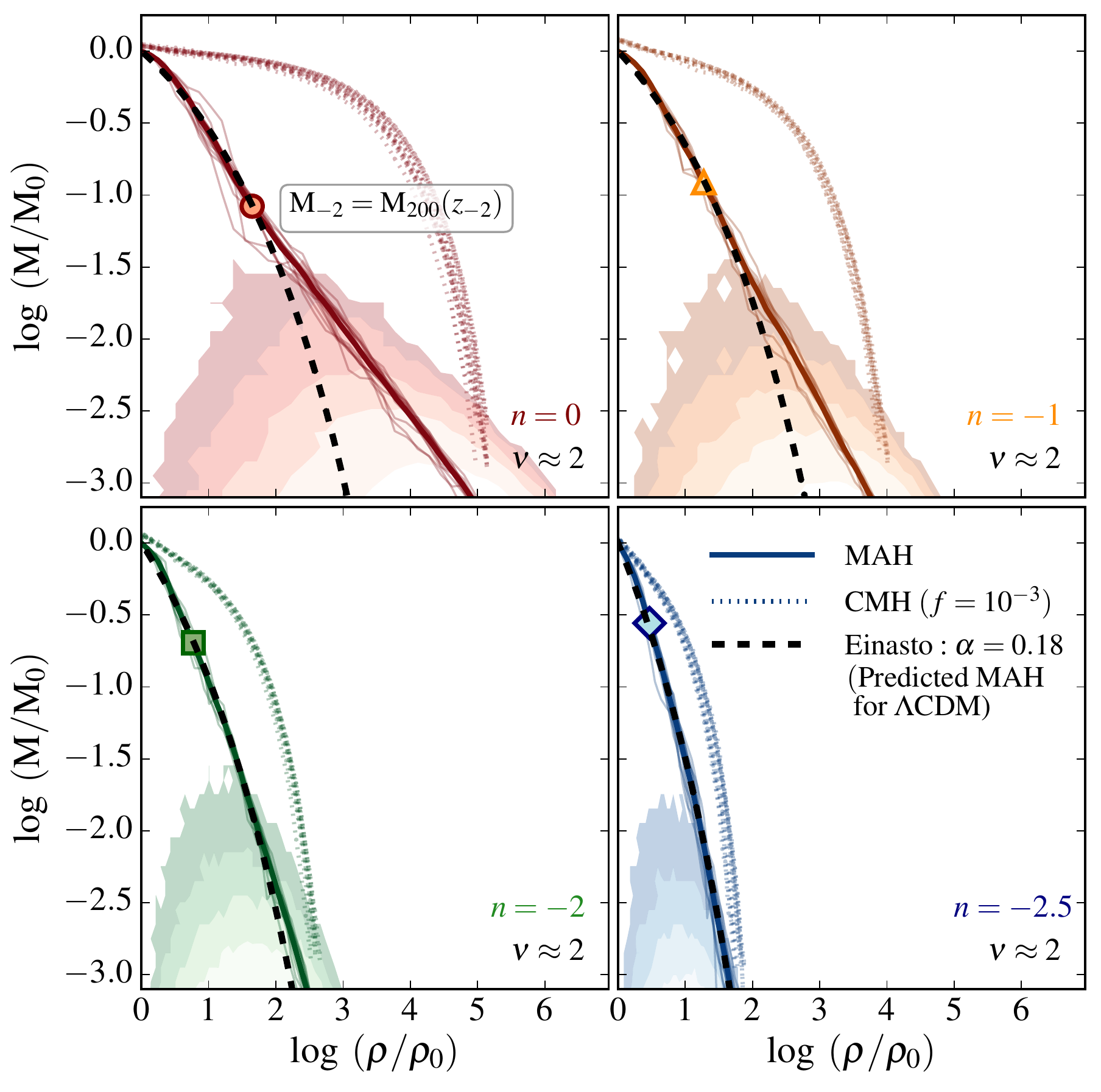}
  \caption{Growth histories of $\nu\approx 2$ halos in our scale-free
    simulations. Thin solid lines show the main progenitor MAHs
    measured at separate redshifts; the thick solid line shows the
    median of these curves. Dotted lines corrsepond the total
    collapsed mass history (CMF) in {\em all} progenitors larger 
    than a fraction $f=10^{-3}$ of the present-day halo mass. Shaded contours
    show the average progenitor mass functions and enclose, respectively,
    1, 2, 5, 10 and 25 progenitors. For comparison, the dashed black line
    is an $\alpha=0.18$ Einasto profile (the ``predicted'' MAH for LCDM
    halos with the same formation time, $z_{-2}$). In all cases masses are normalized
    by the present-day halo mass, ${\rm M}/{\rm M_0}$ and time is expressed 
    as critical density in present-day units, $\rho_c(a)/\rho_0=(a/a_0)^3$. }
  \label{fig:MAH}
\end{figure}

Figure~\ref{fig:MAH} shows these results quantitatively. Here we plot the 
{\em median} MAHs and CMHs of relaxed halos that lie in a narrow range of peak height
($\log \nu=0.3\pm0.05$), separating models of different spectral index into different
panels. Note that we have used the critical 
density, $\rho_{\rm crit}\propto a^3$, as the time variable rather than expansion factor or 
redshift, and have normalized masses and densities to their present-day values, $M_0$ 
and $\rho_0$, respectively. For clarity, results at specific redshifts are shown as thin lines, 
and their average as a thick curve. Whether judged by the MAH or CMH, halos in cosmologies
with larger $n$ collapse earlier, explaining why, e.g., \citet{Knollmann2008}, report
systematically higher concentrations for such systems.

The CMHs betray the fact that, as $n$ decreases, diffuse accretion plays a more
prominent role in halo growth. On average, when each halo's main progenitor first
reached just $\sim 1\%$ of its final mass, the {\em total} collapsed mass fractions 
were $\sim 50\%$, $36\%$, $16\%$ and  $10\%$ for $n=0$, $-1$, $-2$ and $-2.5$, 
respectively. Although smaller $n$ implies larger fractions of diffuse accretion,
merging remains significant in all model, though more so for larger $n$. The shaded 
contours in Figure~\ref{fig:MassiveEvol}, for example, show the full progenitor 
mass functions of these halos (different levels enclose 1, 2, 5 , 10 and 25 progenitors).

Like the CMHs, the median MAHs are approximately self-similar, but their 
{\em shapes} depend strongly on the spectral index, $n$. As expected from 
Figure~\ref{fig:MassiveEvol}, halos in the $n=-2.5$ model grow rapidly,
increasing their virial mass by a factor of $\sim 1000$ over just a factor of $\sim 4$
in expansion history. This is roughly an order of magnitude less than what is required 
for halos the $n=0$ model to grow by the same amount.

Intriguingly, the halo MAHs in the $n=-2$ and $-2.5$ models are
similar to those of $\Lambda$CDM halos. This is not unexpected: the
{\em local slope} of the CDM power spectrum roughly spans $\sim$$-1.8$ to $\sim$$-2.5$ 
for halo masses ranging dwarfs to rich clusters, the mass scale over which
MAHs are well-studied in $\Lambda$CDM models.
The dashed black lines in each panel, for example, 
show an $\alpha=0.18$ Einasto profile 
(expressed here as mass-versus-enclosed density) with the same characteristic
``formation time'', $z_{-2}$, as the scale free MAHs. 
This timescale marks the point at which
the main progneitor's virial mass was first equal to the mass enclosed by its
present-day scale radius, $r_{-2}$ (highlighted as an outsized point in each panel).
This single point can be used to accurately predict the MAHs of $\Lambda$CDM
\citep[see][for a full discussion]{Ludlow2013,Ludlow2016}. 
Note, however, that the slow growth of halos in the $n=0$ and $-1$ models yeild quite 
distinct MAHs from those expected for $\Lambda$CDM. In particular, they are
substantially less curved. Do the diversity of MAHs in scale-free models leave a 
residual imprint on their density profiles?

%__________________________________________
%__________________________________________

\vspace{-0.5cm}
\section{Density Profiles}
\label{SecRho}

\begin{figure*}
  \subfloat{\includegraphics[width=0.50\textwidth]{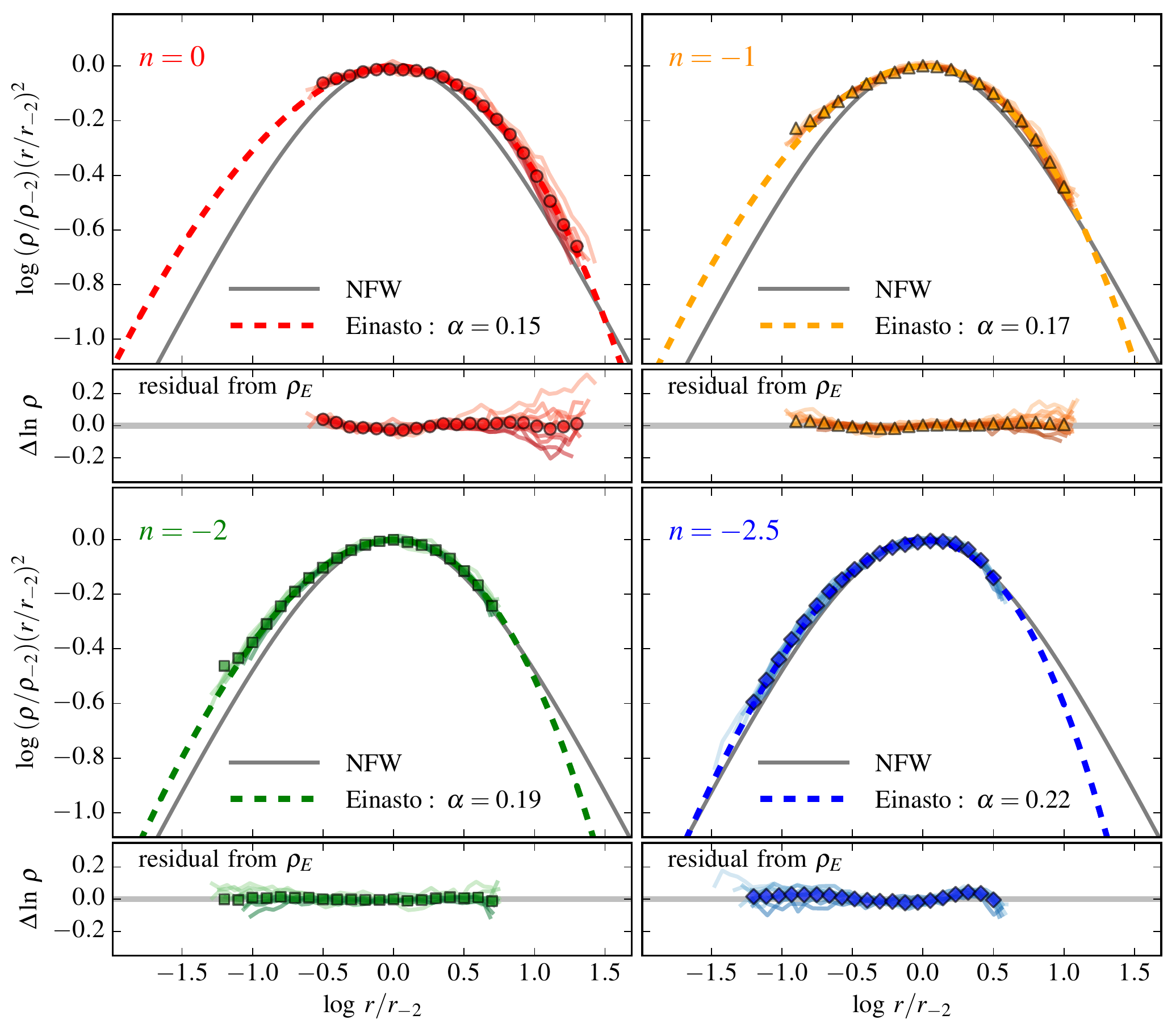}}
  \subfloat{\includegraphics[width=0.23\textwidth]{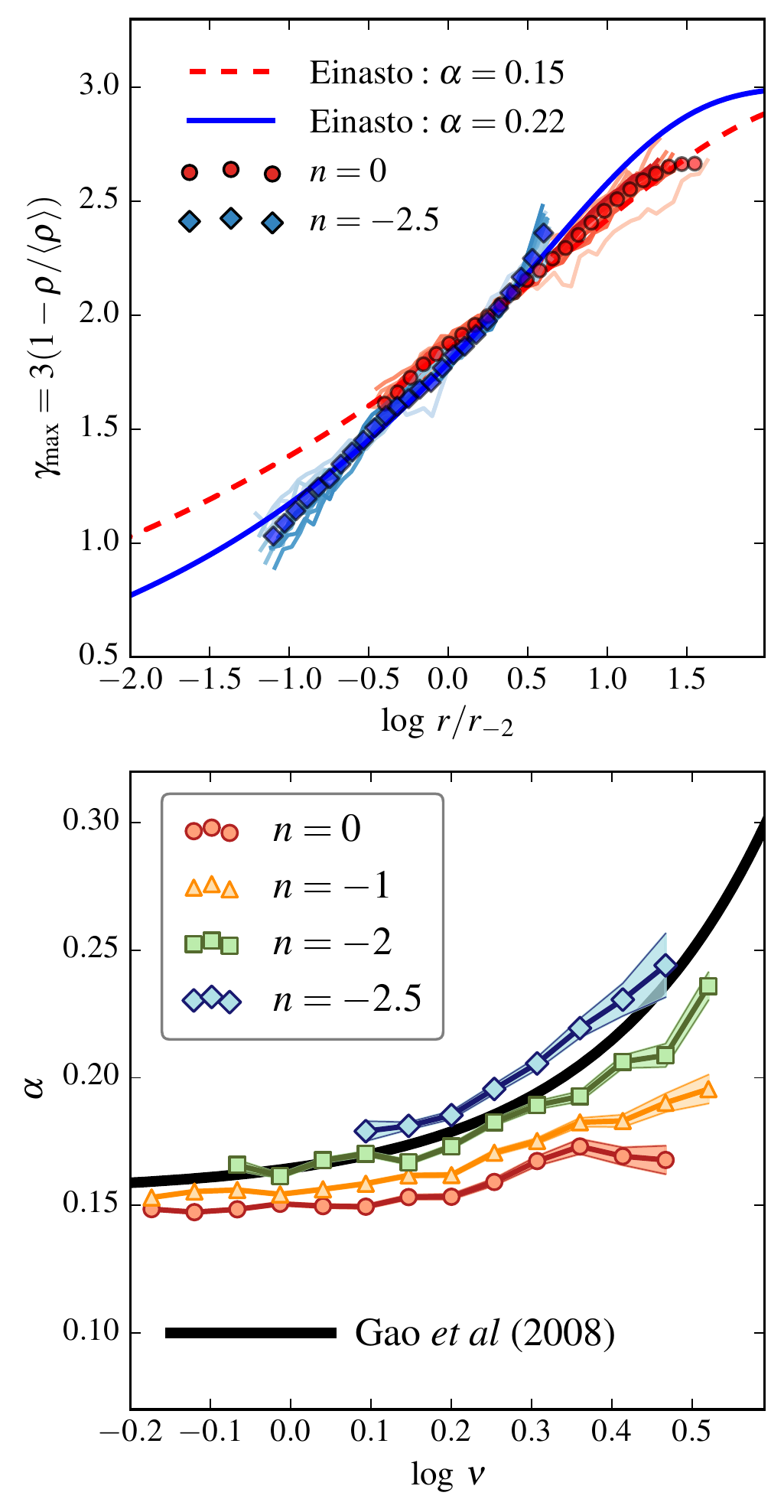}}
  \caption{A summary of the spherically-averaged density profiles of halos identified 
    in our scale-free simulations. The plot on the left, sub-divided in four panels,
    shows the {\em median} density profiles of $\nu\approx 2$ halos at a variety of
    redshift (thin lines). The average of these curves is emphasized using 
    symbols. All profiles have been rescaled by their characteristic density and 
    radius, and weighted by $r^2$.
    For comparison, heavy grey lines show an NFW profile; dashed lines show an 
    Einasto profile whose shape parameter, $\alpha$, was chosen to match the 
    simulated profiles (it is not a best-fit to the data). Residuals in the lower 
    panel are calculated with respect to this profile. The panel on the top-right show
    the maximum asymptotic slope, $\gamma_{\rm max}$, for $n=0$ and $n=-2.5$. The
    median best-fit $\alpha-\nu$ relation {\em all} halos resolved with $\geq 5\times 10^4$ 
    particles is shown in the lower right hand panel.}
  \label{fig:densprof}
\end{figure*}

The left panel of Figure~\ref{fig:densprof} shows
the median spherically-averaged density profiles for the same halos
whose MAHs were plotted in Figure~\ref{fig:MAH}.
To aid the comparison all profiles
are normalized by their characteristic values of density, $\rho_{-2}$,
and radius, $r_{-2}$, and weighted by a factor of $(r/r_{-2})^2$
to enhance dynamic range. 
As above, results from our four simulations are shown in separate panels, using
different colors. Within each panel thin lines (barely distinguishable here)
correspond to different redshifts. Their median is shown using symbols. 

For comparison, we also plot an NFW profile in each panel using a
thick grey line. This curve matches the simulated profiles
reasonably well, even for the white noise $n=0$ model. Nonetheless, important 
differences are also clear. Dashed lines, 
for example, show an Einasto profile whose $\alpha$
was chosen to match that of the simulated halos. For a given $n$ (and
$\nu$) the density profiles are clearly self-similar, regardless of $z$. The residuals 
(lower sub-panels), for example, have been computed with respect to these
Einasto profiles and {\rm are not} deviations from
individual best-fit models. The deviations are not systematic and,
at most radii, remain smaller than $\sim$5\%.

More importantly, the halo density profiles {\em are not} self-similar
across different simulations, even when $\nu$ is held fixed. Instead, $\alpha$ varies 
from $\approx 0.15$ for $n=0$ to $\approx 0.22$ for $n=-2.5$. These differences are
emphasized in the upper-right panel of Figure~\ref{fig:densprof}, where we plot
the maximum asymptotic power-law slope, $\gamma_{\rm max}$, compatible with the inferred mass profiles
(for clarity, we have only included the $n=0$ and $n=-2.5$ runs in this panel).
Clearly, an Einasto profile with a single value of $\alpha$ cannot fit $\gamma_{\rm max}(r)$
for all models simultaneously.

In the lower right-hand panel we show that the power spectrum-dependence of $\alpha$
extends to {\em all} values of $\nu$. Here we plot the best-fit $\alpha-\nu$ relation
obtained for individual halos after combining all redshifts (shaded regions indicate the error on the median values).
As with other panels, different colors and 
symbols correspond to the different simulations. 
{\em At all overlapping $\nu$, the shapes of CDM halo density profiles depend systematically 
on the power spectral index, $n$.}

For comparison, the heavy black line in Figure~\ref{fig:densprof} shows the 
$\alpha-\nu$ relation obtained by \cite[][see also Dutton \& Macci\`{o}, 2014]{Gao2008} 
from the Millennium simulation \citep{Springel2005a}. This curve matches the 
results for our $n=-2$ and $-2.5$ models fairly well, but becomes progressively worse as $n$ 
increases. In particular, halos in our $n=0$ and $-1$ models have, on average, {\em less
curved} mass profiles than those of $\Lambda$CDM halos of similar $\nu$, which was precisely 
the case for their MAHs plotted in Figure~\ref{fig:MAH}.

\vspace{-0.5cm}
\section{Discussion and Conclusions}
\label{SecConc}

Overall, our results imply that {\em the spherically-averaged density profiles
of DM haloes are not universal} but depend systematically on 
the shape of the DM power spectrum. Haloes that grow slowly
through the merger and accretion of many small, dense clumps have
steeper, more centrally concentrated density profiles with extended outer
envelopes. Those that form rapidly, through a combination of diffuse 
accretion and loosely-bound mergers, have shallower
inner profiles and steep outer ones. These results broadly agree with 
the qualitative interpretation put forth by 
\citet[][see also Nipoti 2015]{Cen2014} for the 
origin of Einasto-like density profiles.

The results also support the claim of \citet{Ludlow2013},
who suggested that the curvature of the MAH is what determines
$\alpha$. These authors showed that halos whose MAHs curve more rapidly
than average tend to have more sharply curving density profiles, and
vice versa. Rapid growth implies rapid merging (see Figure~\ref{fig:MAH}), which has
also been shown to enhance the curvature of halo mass profiles \citep{Angulo2016}.

Our results, however, disgree with prior work on halo structure 
in scale-free cosmologies. \citet{Knollmann2008}
found that halo mass profiles are insensitive to differences in the 
fluctuation power spectrum. Although they reported a
positive correlation between the power spectral index, $n$, and the innermost
asymptotic slope $\beta$ of $\rho(r)$, they attributed it to the large
range of halo concentrations spanned in models of widely different $n$,
which lead to difficulties robustly estimating $\beta$.

We hope our results will motivate future studies that 
seek to build a holistic model for halo structure that connects all relevant structural 
parameters to the detailed and unique assembly histories of DM 
halos. Given the complexities involved such a model is unlikely to be simple,
but is within reach of current simulations of halo formation. 

\vspace{-0.75cm}
\section*{acknowledgments}

We thank Alejandro Ben\'{i}tez-Llambay 
for visualization software
(\href{https://github.com/alejandrobll/py-sphviewer}{https://github.com/alejandrobll/py-sphviewer}).
ADL is supported by a COFUND Junior Research Fellowship;
REA by AYA2015-66211-C2-2. This work used the COSMA Data
Centric system at Durham University, operated by the Institute for
Computational Cosmology on behalf of the STFC DiRAC HPC Facility
(www.dirac.ac.uk). This equipment was funded by a BIS National
E-infrastructure capital grant ST/K00042X/1, DiRAC Operations grant
ST/K003267/1 and Durham University. DiRAC is part of the National
E-Infrastructure.

\vspace{-0.5cm}

\bibliographystyle{mn2e}
\bibliography{paper}

\end{document}